\documentstyle[fleqn,11pt]{book}
\begin{document}
\parindent 1.4cm
\large
\begin{center}
{THE QUANTUM WAVE PACKET AND THE FEYNMAN DE BROGLIE BOHM
PROPAGATOR OF THE LINEARIZED SCHUCH CHUNG HARTMANN EQUATION ALONG
A CLASSICAL TRAJETORY}
\end{center}
\begin{center}
{J. M. F. Bassalo$^{1}$,\ P. T. S. Alencar$^{2}$,\  D. G. da
Silva$^{3}$,\ A. B. Nassar$^{4}$\ and\ M. Cattani$^{5}$}
\end{center}
\begin{center}
{$^{1}$\ Funda\c{c}\~ao Minerva,\ Avenida Governador Jos\'e
Malcher\ 629 -\ CEP\ 66035-100,\ Bel\'em,\ Par\'a,\ Brasil
E-mail:\ jmfbassalo@gmail.com}
\end{center}
\begin{center}
{$^{2}$\ Universidade Federal do Par\'a\ -\ CEP\ 66075-900,\
Guam\'a, Bel\'em,\ Par\'a,\ Brasil E-mail:\ tarso@ufpa.br}
\end{center}
\begin{center}
{$^{3}$\ Escola Munguba do Jari, Vit\'oria do Jari\ -\ CEP\
68924-000,\ Amap\'a,\ Brasil E-mail:\ danielgemaque@yahoo.com.br}
\end{center}
\begin{center}
{$^{4}$\ Extension Program-Department of Sciences, University of
California,\ Los Angeles, California 90024,\ USA E-mail:\
nassar@ucla.edu}
\end{center}
\begin{center}
{$^{5}$\ Instituto de F\'{\i}sica da Universidade de S\~ao Paulo.
C. P. 66318, CEP\ 05315-970,\ S\~ao Paulo,\ SP, Brasil E-mail:\
mcattani@if.usp.br}
\end{center}
\par
Abstract:\ In this paper we study the quantum wave
packet and the Feynman-de Broglie-Bohm propagator of the
linearized Schuch-Chung-Hartmann equation along a classical trajetory.
\vspace{0.2cm}
\par
PACS\ 03.65\ -\ Quantum Mechanics
\par
\vspace{0.2cm}
\par
1.\ {\bf Introduction}
\vspace{0.2cm}
\par
In the present work we investigate the quantum wave packet and the
Feynman-de Broglie-Bohm propagator of the linearized Schuch-Chung-Hartmann
equation along a classical trajetory by using the Quantum Mechanical of
the de Broglie-Bohm.$^{[1]}$
\vspace{0.2cm}
\par
2.\ {\bf The Schuch-Chung-Hartmann Equation}
\par
In 1983-1985,$^{[2]}$ D. Schuch, K. M. Chung and H. Hartmann proposed a
non-linear Schr\"{o}dinger to represent time dependent physical
systems, given by:
\begin{center}
{$i\ {\hbar}\ {\frac {{\partial}}{{\partial}t}}\ {\psi}(x,\ t)\ =\ -\
{\frac {{\hbar}^{2}}{2\ m}}\ {\frac {{\partial}^{2}{\psi}(x,\
t)}{{\partial}x^{2}}}\ +\ {\Big {(}}\ V(x,\ t)\ +$}
\end{center}
\begin{center}
{$+\ {\frac {{\hbar}\ {\nu}}{i}}\ [{\ell}n\ {\psi}(x,\ t)\ -\ <{\ell}n\
{\psi}(x,\ t)>]\ {\Big {)}}\ {\psi}(x,\ t)$\ ,\ \ \ \ \ (2.1)}
\end{center}
where ${\psi}(x,\ t)$ and $V(x,\ t)$ are, respectively, the
wavefunction and the time dependent potential of the physical system in
study, and ${\nu}$ is a constant.
\par
Writting the wavefunction ${\psi}(x,\ t)$ in the polar form, defined by
the Madelung-Bohm [3,\ 4]:
\begin{center}
{${\psi}(x,\ t)\ =\ {\phi}(x,\ t)\ exp\ [i\ S(x,\ t)]$,\ \ \ \ \ (2.2)}
\end{center}
where $S(x,\ t)$ is the classical action and ${\phi}(x,\ t)$ will be
defined in what follows, and using eq. (2.2) into eq. (2.1), we get
(remember that ${\ell}n\ e^{i\ S}\ =\ i\ S)$:\ [1]
\par
\begin{center}
{$i\ {\hbar}\ (i\ {\frac {{\partial}S}{{\partial}t}}\ +\ {\frac
{1}{{\phi}}}\ {\frac {{\partial}{\phi}}{{\partial}t}})\ {\psi}\ =$}
\end{center}
\begin{center}
{$=\ -\ {\frac {{\hbar}^{2}}{2\ m}}\ [i\ {\frac
{{\partial}^{2}S}{{\partial}x^{2}}}\ +\ {\frac {1}{{\phi}}}\ {\frac
{{\partial}^{2}{\phi}}{{\partial}x^{2}}}\ -\ ({\frac
{{\partial}S}{{\partial}x}})^{2}\ +\ 2\ {\frac {i}{{\phi}}}\ {\frac
{{\partial}S}{{\partial}x}}\ {\frac {{\partial}{\phi}}{{\partial}x}}]\
{\psi}\ +$}
\end{center}
\begin{center}
{$+\ {\Big {(}}\ V(x,\ t)\ +\ {\frac {{\hbar}\ {\nu}}{i}}\ [{\ell}n\
({\phi}\ e^{i\ S})\ -\ <{\ell}n\ ({\phi}\ e^{i\ S})>]\ {\Big {)}}\
{\psi}\ \ \ {\to}$}
\end{center}
\begin{center}
{$i\ {\hbar}\ (i\ {\frac {{\partial}S}{{\partial}t}}\ +\ {\frac
{1}{{\phi}}}\ {\frac {{\partial}{\phi}}{{\partial}t}})\ {\psi}\ =\ -\
{\frac {{\hbar}^{2}}{2\ m}}\ [i\ {\frac
{{\partial}^{2}S}{{\partial}x^{2}}}\ +\ {\frac {1}{{\phi}}}\ {\frac
{{\partial}^{2}{\phi}}{{\partial}x^{2}}}\ -$}
\end{center}
\begin{center}
{$-\ ({\frac
{{\partial}S}{{\partial}x}})^{2}\ +\ 2\ {\frac {i}{{\phi}}}\ {\frac
{{\partial}S}{{\partial}x}}\ {\frac {{\partial}{\phi}}{{\partial}x}}]\
{\psi}$\ +}
\end{center}
\begin{center}
{$+\ {\Big {(}}\ V(x,\ t)\ -\ i\ {\hbar}\ {\nu}\ [{\ell}n\
{\phi}\ +\ i\ S\ -\ <{\ell}n\ {\phi}>\ -\ i\ <S>]\ {\Big {)}}\ {\psi}$\
.\ \ \ \ \ (2.3)}
\end{center}
\par
Taking the real and imaginary parts of eq. (2.3), we obtain:
\par
a)\ {\underline {imaginary part}}
\begin{center}
{${\frac {1}{{\phi}}}\ {\frac {{\partial}{\phi}}{{\partial}t}}\
=\ -\ {\frac {{\hbar}}{2\ m}}\ ({\frac
{{\partial}^{2}S}{{\partial}x^{2}}}\ +\ {\frac {2}{{\phi}}}\ {\frac
{{\partial}S}{{\partial}x}}\ {\frac {{\partial}{\phi}}{{\partial}x}})\
\ \ -$}
\end{center}
\begin{center}
{$-\ {\nu}\ ({\ell}n\ {\phi}\ -\ <{\ell}n\ {\phi}>)$.\ \ \ \ \
(2.4)}
\end{center}
\par
b)\ {\underline {real part}}
\begin{center}
{$-\ {\hbar}\ {\frac {{\partial}S}{{\partial}t}}\ =\ -\ {\frac
{{\hbar}^{2}}{2\ m}}\ [{\frac {1}{{\phi}}}\ {\frac
{{\partial}^{2}{\phi}}{{\partial}x^{2}}}\ -\ ({\frac
{{\partial}S}{{\partial}x}})^{2}]\ +$}
\end{center}
\begin{center}
{$+\ V(x,\ t)\ +\ {\hbar}\ {\nu}\ (S\ -\ <S>)$. \ \ \ \ \ (2.5)}
\end{center}
\par
\vspace{0.2cm}
2.1 {\bf Dynamics of the Schuch-Chung-Hartmann Equation}
\par
Now, let us to study the dynamics of the Schuch-Chung-Hartmann
equation. To do is let us perform the following correspondences:${[5]}$
\begin{center}
{${\rho}(x,\ t)\ =\ {\phi}^{2}(x,\ t)$\ ,\ \ \ \ \ (2.6)\ \ \
(quantum mass density)}
\end{center}
\begin{center}
{$v_{qu}(x,\ t)\ =\ {\frac {{\hbar}}{m}}\ {\frac {{\partial}S(x,\
t)}{{\partial}x}}$\ ,\ \ \ \ \ (2.7)\ \ \ \ \ (quantum velocity)}
\end{center}
\begin{center}
{$V_{qu}(x,\ t)\ =\ -\ {\frac {{\hbar}^{2}}{2\ m}}\ {\frac {1}{{\sqrt
{{\rho}}}}}\ {\frac {{\partial}^{2}{\sqrt
{{\rho}}}}{{\partial}x^{2}}}\ =\ -\ {\frac {{\hbar}^{2}}{2\ m\
{\phi}}}\ {\frac {{\partial}^{2}{\phi}}{{\partial}x^{2}}}$\
.\ \ \ \ \ (2.8a,b)\ \ \ \ \ (Bohm quantum potential)}
\end{center}
\par
Putting the eqs. (2.6,7) into eq. (2.4) we get [remember that ${\frac
{{\partial}}{{\partial}v}}\ ({\ell}n\ u)\ =\ {\frac {1}{u}}\ {\frac
{{\partial}u}{{\partial}v}}$\ and\ ${\ell}n\ (u^{n})\ =\ n\ {\ell}n\
u$]:
\begin{center}
{${\frac {{\partial}}{{\partial}t}}\ (2\ {\ell}n\ {\phi})\ =$}
\end{center}
\begin{center}
{$=\ -\ {\frac {{\hbar}}{m}}\ [{\frac
{{\partial}^{2}S}{{\partial}x^{2}}}\ +\ {\frac
{{\partial}S}{{\partial}x}}\ {\frac {{\partial}}{{\partial}x}}\ (2\
{\ell}n\ {\phi})]\ -\ 2\ {\nu}\ ({\ell}n\ {\phi}\ -\ <{\ell}n\ {\phi}>)\
\ \ {\to}$}
\end{center}
\begin{center}
{${\frac {{\partial}}{{\partial}t}}\ ({\ell}n\ {\phi}^{2})\ =$}
\end{center}
\begin{center}
{$=\ -\ {\frac {{\hbar}}{m}}\ [{\frac
{{\partial}^{2}S}{{\partial}x^{2}}}\ +\ {\frac
{{\partial}S}{{\partial}x}}\ {\frac {{\partial}}{{\partial}x}}\
({\ell}n\ {\phi}^{2})]\ -\ 2\ {\nu}\ ({\ell}n\ {\phi}\ -\ <{\ell}n\
{\phi}>)\ \ \ {\to}$}
\end{center}
\begin{center}
{${\frac {{\partial}}{{\partial}t}}\ ({\ell}n\ {\rho})\ =$}
\end{center}
\begin{center}
{$=\ -\ {\frac {{\hbar}}{m}}\ [{\frac
{{\partial}^{2}S}{{\partial}x^{2}}}\ +\ {\frac
{{\partial}S}{{\partial}x}}\ {\frac {{\partial}}{{\partial}x}}\
({\ell}n\ {\rho})]\ -\ 2\ {\nu}\ ({\ell}n\ {\sqrt {{\rho}}}\ -\
<{\ell}n\ {\sqrt {{\rho}}}>)\ \ \ {\to}$}
\end{center}
\begin{center}
{${\frac {1}{{\rho}}}\ {\frac {{\partial}{\rho}}{{\partial}t}}\ =$}
\end{center}
\begin{center}
{$=\ -\ {\frac {{\hbar}}{m}}\ ({\frac
{{\partial}^{2}S}{{\partial}x^{2}}}\ +\ {\frac
{{\partial}S}{{\partial}x}}\ {\frac {1}{{\rho}}}\ {\frac
{{\partial}{\rho}}{{\partial}x}})\ -\ {\nu}\ ({\ell}n\ {\rho}\ -\
<{\ell}n\ {\rho}>)\ =$}
\end{center}
\begin{center}
{$=\ -\ {\frac {{\partial}}{{\partial}x}}\ [{\frac {{\hbar}}{m}}\ ({\frac
{{\partial}S}{{\partial}x}})]\ -\ {\frac {1}{{\rho}}}\ {\frac
{{\partial}\ {\rho}}{{\partial}x}}\ [{\frac {{\hbar}}{m}}\ ({\frac
{{\partial}S}{{\partial}x}})]\ -\ {\nu}\ ({\ell}n\ {\rho}\ -\ <{\ell}n\
{\rho}>)\ \ \ {\to}$}
\end{center}
\begin{center}
{${\frac {1}{{\rho}}}\ {\frac {{\partial}{\rho}}{{\partial}t}}\ +\
{\frac {{\partial}v_{qu}}{{\partial}x}}\ +\ {\frac {v_{qu}}{{\rho}}}\
{\frac {{\partial}{\rho}}{{\partial}x}}\ =\ -\ {\nu}\ ({\ell}n\ {\rho}\
-\ <{\ell}n\ {\rho}>)\ \ \ {\to}$}
\end{center}
\begin{center}
{${\frac {{\partial}{\rho}}{{\partial}t}}\ +\ {\rho}\ {\frac
{{\partial}v_{qu}}{{\partial}x}}\ +\ v_{qu}\ {\frac
{{\partial}{\rho}}{{\partial}x}}\ =\ -\ {\nu}\ {\rho}\ ({\ell}n\
{\rho}\ -\ <{\ell}n\ {\rho}>)\ \ \ {\to}$}
\end{center}
\begin{center}
{${\frac {{\partial}{\rho}}{{\partial}t}}\ +\ {\frac {{\partial}({\rho}\
v_{qu})}{{\partial}x}}\ =\ -\ {\nu}\ {\rho}\ ({\ell}n\ {\rho}\ -\
<{\ell}n\ {\rho}>)$\ .\ \ \ \ \ (2.9)}
\end{center}
\par
We must note that the presence of the second member in expression (2.9),
indicates {\underline {descoerence}} of the considered physical system
represented by (2.1).
\par
Now, taking the eq. (2.5) and using the eqs. (2.7,8b), will be:
\begin{center}
{$-\ {\hbar}\ {\frac {{\partial}S}{{\partial}t}}\ =\ -\ ({\frac
{{\hbar}^{2}}{2\ m\ {\phi}}})\ {\frac
{{\partial}^{2}{\phi}}{{\partial}x^{2}}}\ +\ {\frac {1}{2}}\ m\ ({\frac
{{\hbar}}{m}}\ {\frac {{\partial}S}{{\partial}x}})^{2}\ +$}
\end{center}
\begin{center}
{$+\ V(x,\ t)\ +\ {\hbar}\ {\nu}\ (S\ - <S>)\ \ \ {\to}$}
\end{center}
\begin{center}
{${\hbar}\ [{\frac {{\partial}S}{{\partial}t}}\ +\ {\nu}\ (S\ -\ <S>)]\ +\
{\frac {1}{2}}\ m\ v_{qu}^{2}\ +\ V\ +\ V_{qu}\ =\ 0$\ .\ \ \ \ \
(2.10)}
\end{center}
\par
Considering that:
\begin{center}
{$<f(x,\ t)>\ =\ {\int}_{-\ {\infty}}^{+\ {\infty}}\ {\rho}(x,\ t)\
f(x,\ t) dx\ =\ g(t)$\ ,\ \ \ \ \ (2.11)}
\end{center}
then:
\begin{center}
{${\frac {{\partial}<S>}{{\partial}x}}\ =\ {\frac
{{\partial}}{{\partial}x}}\ {\int}_{-\ {\infty}}^{+\ {\infty}}\
{\rho}(x,\ t)\ S(x,\ t) dx\ =\ {\frac {{\partial}g(t)}{{\partial}x}}\
=\ 0$\ .\ \ \ \ \ (2.12)}
\end{center}
\par
Now, differentiating the eq. (2.5) with respect $x$, and using the eqs.
(2.7,8b,12), we have:
\begin{center}
{$-\ {\hbar}\ {\frac {{\partial}^{2}S}{{\partial}x\
{\partial}t}}\ =\ -\ {\frac {{\hbar}^{2}}{2\ m}}\ {\frac
{{\partial}}{{\partial}x}}\ [{\frac {1}{{\phi}}}\ {\frac
{{\partial}^{2}{\phi}}{{\partial}x^{2}}}\ -\ ({\frac
{{\partial}S}{{\partial}x}})^{2}]\ +\ {\frac
{{\partial}V}{{\partial}x}}\ +\ {\hbar}\ {\nu}\ ({\frac
{{\partial}S}{{\partial}x}} -\ {\frac {{\partial}<S>}{{\partial}x}})\ \
\ {\to}$}
\end{center}
\begin{center}
{$-\ {\frac {{\partial}}{{\partial}t}}\ ({\frac {{\hbar}}{m}}\ {\frac
{{\partial}S}{{\partial}x}})\ =\ {\frac {1}{m}}\ {\frac
{{\partial}}{{\partial}x}}\ (-\ {\frac {{\hbar}^{2}}{2\ m}}\ {\frac
{1}{{\phi}}}\ {\frac {{\partial}^{2}{\phi}}{{\partial}x^{2}}})$\ +}
\end{center}
\begin{center}
{+\ ${\frac {1}{2}}\ {\frac {{\partial}}{{\partial}x}}\ ({\frac
{{\hbar}}{m}}\ {\frac {{\partial}S}{{\partial}x}})^{2}\ +\
{\frac {1}{m}}\ {\frac {{\partial}V}{{\partial}x}}\ +\ {\nu}\ {\frac
{{\hbar}}{m}}\ {\frac {{\partial}S}{{\partial}x}} -\ {\frac
{{\hbar}}{m}}\ {\frac {{\partial}<S>}{{\partial}x}}\ \ \ {\to}$}
\end{center}
\begin{center}
{${\frac {{\partial}v_{qu}}{{\partial}t}}\ +\ v_{qu}\ {\frac
{{\partial}v_{qu}}{{\partial}x}}\ +$}
\end{center}
\begin{center}
{$+\ {\nu}\ v_{qu}\ = -\ {\frac {1}{m}}\ {\frac
{{\partial}}{{\partial}x}}\ (V\ +\ V_{qu})$\ .\ \ \ \ \ (2.13)}
\end{center}
\par
Considering the "substantive differentiation" (local plus convective)
or "hidrodynamic differention":\ $d/dt\ =\ {\partial}/{\partial}t\ +
v_{qu}\ {\partial}/{\partial}x$ and that $v_{qu}\ =\ dx_{qu}/dt$, the
eq. (2.13) could be written as:[5]
\begin{center}
{$m\ d^{2}x/dt^{2}\ =\ -\ {\nu}\ v_{qu}\ -\ {\frac {1}{m}}\ {\frac
{{\partial}}{{\partial}x}}\ (V\ +\ V_{qu})$\ ,\ \ \ \ \ (2.14)}
\end{center}
that has a form of the $Second\ Newton\ Law$.
\vspace{0.2cm}
\par
3.\ {\bf The Quantum Wave Packet of the Linearized Schuch-Chung-Hartmann
Equation along a Classical Trajetory}
\vspace{0.2cm}
\par
In order to find the quantum wave packet of the Schuch-Chung-Hartmann
equation, let us considerer the following $ansatz$:[6]
\par
\begin{center}
{${\rho}\ (x,\ t) =\ [2{\pi}\ a^{2}(t)]^{-\ 1/2}\ exp\ {\Big {(}}\ -\
{\frac {[x\ -\ q(t)]^{2}}{2\ a^{2}(t)}}\ {\Big {)}}$,\ \ \ \ \ (3.1)}
\end{center}
where $a(t)$ and $q(t)$ are auxiliary functions of time, to be
determined in what follows;\ they represent the {\it width} and {\it center
of mass of wave packet}, respectively.
\par
Taking the eq. (3.1), let us calculated the expressions (remember
that ${\ell}n\ e^{{\alpha}}\ =\ {\alpha}$):
\begin{center}
{${\ell}n\ {\rho}\ (x,\ t)\ =\ {\ell}n\ {\Big {(}}\ [2\ {\pi}\
a^{2}(t)]^{-\ 1/2}\ e^{-\ {\frac {[x\ -\ q(t)]^{2}}{2\ a^{2}(t)}}}\
{\Big {)}}\ =$}
\end{center}
\begin{center}
{$=\ {\ell}n\ [2\ {\pi}\ a^{2}(t)]^{-\ 1/2}\ -\ {\frac {[x\ -\
q(t)]^{2}}{2\ a^{2}(t)}}$\ .\ \ \ \ \ (3.2)}
\end{center}
\begin{center}
{$<{\ell}n\ {\rho}\ (x,\ t)>\ =\ <{\ell}n\ {\Big {(}} [2\ {\pi}\
a^{2}(t)]^{-\ 1/2}\ e^{-\ {\frac {[x\ -\ q(t)]^{2}}{2\ a^{2}(t)}}}\
{\Big {)}}>\ =$}
\end{center}
\begin{center}
{$=\ {\ell}n\ [2\ {\pi}\ a^{2}(t)]^{-\ 1/2}\ -\ <{\frac {[x\ -\
q(t)]^{2}}{2\ a^{2}(t)}}>$\ .\ \ \ \ \ (3.3)}
\end{center}
\par
Considering that:
\begin{center}
{${\int}_{-\ {\infty}}^{{\infty}}\ z^{2}\ e^{-\ z^{2}}\ dz\ =\ {\frac
{{\sqrt {{\pi}}}}{2}}$\ ,}
\end{center}
and the eq. (2.11), we have:\ [1]
\begin{center}
{${\ell}n\ {\rho}\ -\ <{\ell}n\ {\rho}>\ =\ -\ {\frac {a^{2}}{2\
{\rho}}}\ {\frac {{\partial}^{2}{\rho}}{{\partial}x^{2}}}$\ .\ \ \ \ \
(3.4)}
\end{center}
\par
Insering the eq. (3.4) into eq. (2.9), results:
\begin{center}
{${\frac {{\partial}{\rho}}{{\partial}t}}\ +\ {\frac {{\partial}({\rho}\
v_{qu})}{{\partial}x}}\ =\ -\ {\nu}\ {\rho}\ (-\ {\frac {a^{2}}{2\
{\rho}}}\ {\frac {{\partial}^{2}{\rho}}{{\partial}x^{2}}})\ \ \ {\to}$}
\end{center}
\begin{center}
{${\frac {{\partial}{\rho}}{{\partial}t}}\ +\ {\frac {{\partial}({\rho}\
v_{qu})}{{\partial}x}}\ -\ {\frac {{\partial}}{{\partial}x}}\ (D\ {\frac
{{\partial}{\rho}}{{\partial}x}})\ =\ 0$,\ \ \ \ \ (3.5a)}
\end{center}
where:
\begin{center}
{$D\ =\ {\frac {{\nu}\ a^{2}}{2}}$\ .\ \ \ \ \ (3.5b)}
\end{center}
\par
Defining:\ [7]
\begin{center}
{${\vartheta}_{qu}\ =\ v_{qu}\ -\ {\frac {D}{{\rho}}}\ {\frac
{{\partial}{\rho}}{{\partial}x}}$\ ,\ \ \ \ \ (3.6a)}
\end{center}
then the eq. (3.5a) will be the form:
\begin{center}
{${\frac {{\partial}{\rho}}{{\partial}t}}\ +\ {\frac {{\partial}({\rho}\
{\vartheta}_{qu})}{{\partial}x}}\ =\ 0$.\ \ \ \ \ (3.6b)}
\end{center}
\par
Now, substituting (3.1) into (3.5a) and integrated the result, we
obtain:\ [1]
\begin{center}
{${\vartheta}_{qu}(x,\ t)\ =\ {\frac {{\dot {a}}(t)}{a(t)}}\ [x\ -\
q(t)]\ +\ {\dot {q}}(t)$\ ,\ \ \ \ \ (3.7a)}
\end{center}
and:
\begin{center}
{$v_{qu}\ (x,\ t)\ =\ [{\frac {{\dot
{a}}(t)}{a(t)}}\ -\ {\frac {{\nu}}{2}}]\ [x\ -\ q(t)]\ +\ {\dot {q}}(t)$\
.\ \ \ \ \ (3.7b)}
\end{center}
\par
To obtain the quantum wave packet of the linearized
Schuch-Chung-Hartmann equation along a classical trajetory given by
(2.1), let us expand the functions $S(x,\ t)$,\ \ $V(x,\ t)$ and
$V_{qu}(x,\ t)$ around of $q(t)$ up to second Taylor order.\ In this
way we have:
\begin{center}
{$S(x,\ t)\ =\ S[q(t),\ t]\ +\ S'[q(t),\ t]\ [x\ -\ q(t)]\ +\ {\frac
{S''[q(t),\ t]}{2}}\ [x\ -\ q(t)]^{2}$\ ,\ \ \ \ \ (3.8)}
\end{center}
\begin{center}
{$V(x,\ t)\ =\ V[q(t),\ t]\ +\ V'[q(t),\ t]\ [x\ -\
q(t)]\ +\ {\frac {V''[q(t),\ t]}{2}}\ [x\ -\ q(t)]^{2}$\ .\ \ \ \ \ (3.9)}
\end{center}
\begin{center}
{$V_{qu}(x,\ t)\ =\ V_{qu}[q(t),\ t]\ +\ V_{qu}'[q(t),\ t]\ [x\ -\
q(t)]\ +\ {\frac {V_{qu}''[q(t),\ t]}{2}}\ [x\ -\ q(t)]^{2}$\ .\ \ \ \
\ (3.10)}
\end{center}
\par
Differentiating (3.8) in the variable $x$, multiplying
the result by ${\frac {{\hbar}}{m}}$, using the eqs. (2.7) and
(3.7b), taking into account the polynomial identity property and also
considering the second Taylor order, we obtain:
\begin{center}
{${\frac {{\hbar}}{m}}\ {\frac {{\partial}S(x,\ t)}{{\partial}x}}\ =\
{\frac {{\hbar}}{m}}\ {\Big {(}}\ S'[q(t),\ t]\ +\ S''[q(t),\ t]\ [x\
-\ q(t)]\ {\Big {)}}\ =$}
\end{center}
\begin{center}
{$=\ v_{qu}(x,\ t)\ =\ [{\frac {{\dot
{a}}(t)}{a(t)}}\ -\ {\frac {{\nu}}{2}}]\ [x\ -\ q(t)]\ +\ {\dot {q}}(t)\
\ \ \ {\to}$}
\end{center}
\begin{center}
{$S'[q(t),\ t]\ =\ {\frac {m\ {\dot {q}}(t)}{{\hbar}}}\ ,\ \ \
S''[q(t),\ t]\ =\ {\frac {m}{{\hbar}}}\ [{\frac {{\dot
{a}}(t)}{a(t)}}\ -\ {\frac {{\nu}}{2}}]$\ ,\ \ \ \ \ (3.11a,b)}
\end{center}
\par
Substituting (3.11a,b) into (3.8), results:
\begin{center}
{$S(x,\ t)\ =\ S_{o}(t)\ +\ {\frac {m\ {\dot {q}}(t)}{{\hbar}}}\ [x\ -\
q(t)]\ +\ {\frac {m}{2\ {\hbar}}}\ [{\frac {{\dot
{a}}(t)}{a(t)}}\ -\ {\frac {{\nu}}{2}}]\ [x\
-\ q(t)]^{2}$\ ,\ \ \ \ \ (3.12a)}
\end{center}
where:
\begin{center}
{$S_{o}(t)\ {\equiv}\ S[q(t),\ t]$\ ,\ \ \ \ \ (3.12b)}
\end{center}
are the classical actions.
\par
Now, considering that:
\begin{center}
{${\int}_{-\ {\infty}}^{{\infty}}\ z^{n}\ e^{-\ z^{2}}\ dz\ =\ {\frac
{{\sqrt {{\pi}}}}{2}};\ 0;\ {\sqrt {{\pi}}}$\ ,}
\end{center}
respectivelly, for n\ =\ 2,\ 1,\ 0, and using the eqs. (2.11), (3.1)
and (3.12a), we have:
\begin{center}
{$<S>\ =\ {\int}_{-\ {\infty}}^{+\ {\infty}}\ {\rho}(x,\ t)\
S(x,\ t) dx\ =\ S_{1}\ +\ S_{2}\ +\ S_{3}$,\ \ \ \ \ \ (3.13a)}
\end{center}
where:
\begin{center}
{$S_{1}\ =\ {\int}_{-\ {\infty}}^{+\ {\infty}}\ [2{\pi}\ a^{2}(t)]^{-\
1/2}\ exp\ {\Big {(}}\ -\ {\frac {[x\ -\ q(t)]^{2}}{2\ a^{2}(t)}}\
{\Big {)}}\ S_{0}(t)\ dx\ =\ S_{0}(t)$,\ \ \ \ \ (3.13b)}
\end{center}
\begin{center}
{$S_{2}\ =\ {\int}_{-\ {\infty}}^{+\ {\infty}}\ [2{\pi}\ a^{2}(t)]^{-\
1/2}\ exp\ {\Big {(}}\ -\ {\frac {[x\ -\ q(t)]^{2}}{2\ a^{2}(t)}}\
{\Big {)}}\ {\frac {m\ {\dot {q}}(t)}{{\hbar}}}\ [x\ -\
q(t)]\ dx\ =\ 0$,\ \ \ \ \ (3.13c)}
\end{center}
\begin{center}
{$S_{3}\ =\ {\int}_{-\ {\infty}}^{+\ {\infty}}\ [2{\pi}\ a^{2}(t)]^{-\
1/2}\ exp\ {\Big {(}}\ -\ {\frac {[x\ -\ q(t)]^{2}}{2\ a^{2}(t)}}\
{\Big {)}}\ {\frac {m}{2\ {\hbar}}}\ [{\frac {{\dot
{a}}(t)}{a(t)}}\ -\ {\frac {{\nu}}{2}}]\ [x\
-\ q(t)]^{2}\ dx$\ =}
\end{center}
\begin{center}
{$=\ {\frac {m}{2\ {\hbar}}}\ [{\frac {{\dot
{a}}(t)}{a(t)}}\ -\ {\frac {{\nu}}{2}}]$.\ \ \ \ \ (3.13d)}
\end{center}
Inserting the eqs. (3.13b-d) into eq. (3.13a), results:
\begin{center}
{$<S>\ =\ S_{0}(t)\ +\ {\frac {m}{2\ {\hbar}}}\ [{\frac {{\dot
{a}}(t)}{a(t)}}\ -\ {\frac {{\nu}}{2}}]$.\ \ \ \ \ (3.14)}
\end{center}
\par
Differentiating the (3.12a) with respect to $t$, we obtain
(remembering that ${\frac {{\partial}x}{{\partial}t}}$\ =\ 0):
\begin{center}
{${\frac {{\partial}S\ (x,\ t)}{{\partial}t}}\ =\ {\dot {S}}_{o}(t)\ +\
{\frac {{\partial}}{{\partial}t}}\ {\big {[}}\ {\frac {m\ {\dot
{q}}(t)}{{\hbar}}}\ [x\ -\ q(t)]\ {\big {]}}\ +\ {\frac
{{\partial}}{{\partial}t}}\ {\Big {[}}\ {\frac {m}{2\ {\hbar}}}\ {\Big
{(}}\ [{\frac {{\dot {a}}(t)}{a(t)}}\ -\ {\frac {{\nu}}{2}}]\ {\Big
{)}}\ [x\ -\ q(t)]^{2}\ {\Big {]}}\ \ \ {\to}$}
\end{center}
\begin{center}
{${\frac {{\partial}S\ (x,\ t)}{{\partial}t}}\ =\ {\dot {S}}_{o}(t)\ +\
{\frac {m\ {\ddot {q}}(t)}{{\hbar}}}\ [x\ -\ q(t)]\ -\ {\frac {m\ {\dot
{q}}^{2}(t)}{{\hbar}}}\ +$}
\end{center}
\begin{center}
{+\ ${\frac {m}{2\ {\hbar}}}\ [{\frac {{\ddot
{a}}(t)}{a(t)}}\ -\ {\frac {{\dot
{a}}^{2}(t)}{a^{2}(t)}}]\ [x\ -\ q(t)]^{2}\ -\ {\frac {m\
{\dot {q}}(t)}{{\hbar}}}\ {\Big {(}}\ {\frac {{\dot
{a}}(t)}{a(t)}}\ -\ {\frac {{\nu}}{2}}\ {\Big {)}}\ [x\ -\ q(t)]$\ .\ \
\ \ \ (3.15)}
\end{center}
\par
Considering the eqs. (2.6) and (3.1), let us write $V_{qu}$ given by
(2.8a,b) in terms of potencies of $[x\ -\ q(t)]$. Before, we calculate
the following derivations:
\begin{center}
{${\frac {{\partial}{\phi}\ (x,\ t)}{{\partial}x}}\ =\ {\frac
{{\partial}}{{\partial}x}}\ {\Big {(}}\ [2\ {\pi}\ a^{2}(t)]^{-\
1/4}\ e^{-\ {\frac {[x\ -\ q(t)]^{2}}{4\ a^{2}(t)}}}\ {\Big
{)}}\ =\ [2\ {\pi}\ a^{2}(t)]^{-\ 1/4}\ e^{-\ {\frac {[x\ -\
q(t)]^{2}}{4\ a^{2}(t)}}} {\frac {{\partial}}{{\partial}x}}\
{\Big {(}}\ -\ {\frac {[x\ -\ q(t)]^{2}}{4\ a^{2}(t)}}\ {\Big
{)}}\ \ \ {\to}$}
\end{center}
\begin{center}
{${\frac {{\partial}{\phi}\ (x,\ t)}{{\partial}x}}\ =\ -\ [2\ {\pi}\
a^{2}(t)]^{-\ 1/4}\ e^{-\ {\frac {[x\ -\ q(t)]^{2}}{4\
^{2}(t)}}}\ {\frac {[x\ -\ q(t)]}{2\ a^{2}(t)}}$\ ,}
\end{center}
\begin{center}
{${\frac {{\partial}^{2}{\phi}\ (x,\ t)}{{\partial}x^{2}}}\ =\ {\frac
{{\partial}}{{\partial}x}}\ {\Big {(}}\ -\ [2\ {\pi}\
a^{2}(t)]^{-\ 1/4}\ e^{-\ {\frac {[x\ -\ q(t)]^{2}}{4\
a^{2}(t)}}}\ {\frac {[x\ -\ q(t)]}{2\ a^{2}(t)}}\ {\Big
{)}}$\ =}
\end{center}
\begin{center}
{$\ =\ -\ [2\ {\pi}\ a^{2}(t)]^{-\ 1/4}\ e^{-\ {\frac {[x\ -\
q(t)]^{2}}{4\ a^{2}(t)}}}\ {\frac {{\partial}}{{\partial}x}}\
{\Big {(}}\ {\frac {[x\ -\ q(t)]}{2\ a^{2}(t)}}\ {\Big {)}}\ -$}
\end{center}
\begin{center}
{$-\ [2\ {\pi}\ a^{2}(t)]^{-\ 1/4}\ e^{-\ {\frac {[x\ -\
q(t)]^{2}}{4\ a^{2}(t)}}}\ {\frac {[x\ -\ q(t)]}{2\ a^{2}(t)}}\ {\frac
{{\partial}}{{\partial}x}}\ {\Big {(}}\ -\ {\frac {[x\ -\ q(t)]^{2}}{4\
a^{2}(t)}}\ {\Big {)}}\ \ \ {\to}$}
\end{center}
\begin{center}
{${\frac {{\partial}^{2}{\phi}\ (x,\ t)}{{\partial}x^{2}}}\ =\ -\ [2\
{\pi}\ a^{2}(t)]^{-\ 1/4}\ e^{-\ {\frac {[x\ -\ q(t)]^{2}}{4\
a^{2}(t)}}}\ {\frac {1}{2\ a^{2}(t)}}\ +\ [2\ {\pi}\ a^{2}(t)]^{-\
1/4}\ e^{-\ {\frac {[x\ -\ q(t)]^{2}}{4\ a^{2}(t)}}}\ {\frac {[x\ -\
q(t)]^{2}}{4\ a^{4}(t)}}$\ =}
\end{center}
\begin{center}
{$=\ -\ {\phi}\ (x,\ t)\ {\frac {1}{2\ a^{2}(t)}}\ +\ {\phi}\ (x,\ t)\
{\frac {[x\ -\ q(t)]^{2}}{4\ a^{4}(t)}}\ \ \ {\to}$}
\end{center}
\begin{center}
{${\frac {1}{{\phi}\ (x,\ t)}}\ {\frac {{\partial}^{2}{\phi}\ (x,\
t)}{{\partial}x^{2}}}\ =\ {\frac {[x\ -\ q(t)]^{2}}{4\ a^{4}(t)}}\ -\
{\frac {1}{2\ a^{2}(t)}}$\ .\ \ \ \ \ (3.16)}
\end{center}
\par
Substituting (3.16) into (2.8b) and taking into account (3.10), results:
\begin{center}
{$V_{qu}(x,\ t)\ =\ {\frac {{\hbar}^{2}}{4\ m\ a^{2}(t)}}\ -\ {\frac
{{\hbar}^{2}}{8\ m\ a^{4}(t)}}\ [x\ -\ q(t)]^{2}$\ .\ \ \ \ \ (3.17)}
\end{center}
\begin{center}
{$\ V_{qu}[q(t),\ t]\ =\ {\frac {{\hbar}^{2}}{4\ m\ a^{2}(t)}}$\ ,\ \ \
\ \ (3.18a)}
\end{center}
\begin{center}
{$\  V_{qu}'[q(t),\ t]\ =\ 0, \ \ \ V_{qu}''[q(t),\ t]\ =\ -\ {\frac
{{\hbar}^{2}}{4\ m\ a^{4}(t)}}$\ .\ \ \ \ \ (3.18b,c)}
\end{center}
\par
Inserting the eqs. (3.7b,8,9) and (3.12a,14,15,17), into
(2.10), we obtain [remem\-bering that $S_{o}(t)$, $a(t)$ and $q(t)$]:
\begin{center}
{${\hbar}\ [{\frac {{\partial}S}{{\partial}t}}\ +\ {\nu}\ (S\ -\ <S>)]\
+\ {\frac {1}{2}}\ m\ v_{qu}^{2}\ +\ V\ +\ V_{qu}\ =$}
\end{center}
\begin{center}
{$=\ {\hbar}\ {\dot {S}}_{o}(t)\ +\ m\ {\ddot
{q}}(t)\ [x\ -\ q(t)]\ -\ m\ {\dot
{q}}^{2}(t)\ +\ {\frac {m}{2}}\ {\big {[}}\ {\frac
{{\ddot {a}}(t)}{a(t)}}\ -\ {\frac {{\dot
{a}}^{2}(t)}{a^{2}(t)}}\ {\big {]}}\ [x\ -\ q(t)]^{2}\ -$}
\end{center}
\begin{center}
{$-\ m\ {\dot {q}}(t)\ [{\frac {{\dot {a}}(t)}{a(t)}}\ -\ {\frac
{{\nu}}{2}}\ ]\ [x\ -\ q(t)]\ +\ {\nu}\ {\Big {(}}\ {\hbar}\ S_{0}(t)\
+\ m\ {\dot {q}}(t)\ [x\ -\ q(t)]$\ +}
\end{center}
\begin{center}
{$+\ {\frac {m}{2}}\ [{\frac {{\dot  {a}}(t)}{a(t)}}\ -\ {\frac
{{\nu}}{2}}]\ [x\ -\ q(t)]^{2}\ -\ {\hbar}\ S_{0}(t)\ -\ {\frac
{m}{2}}\ [{\frac {{\dot {a}}(t)}{a(t)}}\ -\ {\frac {{\nu}}{2}}]\ {\Big
{)}}\ +$}
\end{center}
\begin{center}
{$+\ {\frac {m}{2}}\ [{\frac{{\dot {a}}(t)}{a(t)}}\ -\
{\frac {{\nu}}{2}}]^{2}\ [x\ -\ q(t)]^{2}\ +\ m\ {\dot {q}}(t)\ [{\frac{{\dot
{a}}(t)}{a(t)}}\ -\ {\frac {{\nu}}{2}}]\ [x\ -\ q(t)]\ +\ {\frac {m\
{\dot {q}}^{2}(t)}{2}}\ +$}
\end{center}
\begin{center}
{$+\ V[q(t),\ t]\ +\ V'[q(t),\ t]\ [x\ -\ q(t)]\ +\ {\frac {1}{2}}\
V''[q(t),\ t]\ [x\ -\ q(t)]^{2}$\ +}
\end{center}
\begin{center}
{$+\ {\frac {{\hbar}^{2}}{4\ m\ a^{2}(t)}}\ -\ {\frac
{{\hbar}^{2}}{8\ m\ a^{4}(t)}}\ [x\ -\ q(t)]^{2}\ =\ 0$\ .\ \ \ \ \
(3.19)}
\end{center}
\par
Expanding the eq. (3.19) in potencies of $[x\ -\ q(t)]$, we obtain
(remember that $[x\ -\ q(t)]^{o}\ =\ 1$):
\begin{center}
{${\Big {(}}\ {\hbar}\ {\dot {S}}_{o}(t)\ -\ {\frac {1}{2}}\ m\ {\dot
{q}}^{2}(t)\ -\ {\frac {m\ {\nu}}{2}}\ [{\frac {{\dot {a}}(t)}{a(t)}}\
-\ {\frac {{\nu}}{2}}]\ +\ V[q(t),\ t]\ +\ {\frac {{\hbar}^{2}}{4\ m\
a^{2}(t)}}\ {\Big {)}}\ [x\ -\ q(t)]^{o}\
+$}
\end{center}
\begin{center}
{$+\ {\Big {(}}\ m\ {\ddot {q}}(t)\ +\ {\nu}\ m\ {\dot {q}}(t)\ +\
V'[q(t),\ t]\ {\Big {)}}\ [x\ -\ q(t)]$\ +}
\end{center}
\begin{center}
{$+\ {\Big {(}}\ {\frac {m}{2}}\ {\frac {{\ddot {a}}(t)}{a(t)}}\ -\ {\frac
{{\nu}^{2}\ m}{8}}\ +\ {\frac {1}{2}}\ V"[q(t),\ t]\ -\ {\frac
{{\hbar}^{2}}{8\ m\ a^{4}(t)}}\ {\Big {)}}\ [x\ -\ q(t)]^{2}\ =\ 0$\ .\
\ \ \ \ (3.22)}
\end{center}
\par
As (3.22) is an identically null polynomium, all coefficients of the
potencies must be all equal to zero, that is:
\begin{center}
{${\dot {S}}_{o}(t)\ =\ {\frac {1}{{\hbar}}} {\Big {(}}\ {\frac
{1}{2}}\ m\ {\dot {q}}^{2}(t)\ +\ {\frac {{\nu}\ m}{2}}\ [{\frac {{\dot
{a}}(t)}{a(t)}}\ -\ {\frac {{\nu}}{2}}]\ -\ V[q(t),\ t]\ -\ {\frac
{{\hbar}^{2}}{4\ m\ a^{2}(t)}}\ {\Big {)}}$\ ,\ \ \ \ \ (3.23)}
\end{center}
\begin{center}
{${\ddot {q}}(t)\ +\ {\nu}\ {\dot {q}}(t)\ +\ {\frac {1}{m}}\ V'[q(t),\
t]\ =\ 0$,\ \ \ \ \ (3.24)}
\end{center}
\begin{center}
{${\ddot {a}}(t)\ +\ a(t)\ {\Big {(}}\ {\frac {1}{m}}\ V"[q(t),\ t]\
-\ {\frac {{\nu}^{2}}{4}}\ {\Big {)}}\ =\ {\frac {{\hbar}^{2}}{4\
m^{2}\ a^{3}(t)}}$\ .\ \ \ \ \ (3.25)}
\end{center}
\par
Assuming that the following initial conditions are obeyed:
\begin{center}
{$q(0)\ =\ x_{o}\ ,\ \ \ {\dot {q}}(0)\ =\ v_{o}\ ,\ \ \ a(0)\
=\ a_{o}\ ,\ \ \ {\dot {a}}(0)\ =\ b_{o}$\ ,\ \ \ \ \ \ (3.26a-d)}
\end{center}
and that:
\begin{center}
{$S_{o}(0)\ =\ {\frac {m\ v_{o}\ x_{o}}{{\hbar}}}$\ ,\ \ \ \ \ (3.27)}
\end{center}
the integration of (3.23) gives:
\begin{center}
{$S_{o}(t)\ =\ {\frac {1}{{\hbar}}}\ {\int}_{o}^{t}\ dt'\ {\Big {(}}\
{\frac {1}{2}}\ m\ {\dot {q}}^{2}(t')\ +\ {\frac {m\ {\nu}}{2}}\
[{\frac {{\dot {a}}(t)}{a(t)}}\ -\ {\frac {{\nu}}{2}}]$\ -}
\end{center}
\begin{center}
{$-\ V[q(t'),\ t']\ -\ {\frac {{\hbar}^{2}}{4\ m\ a^{2}(t')}}\ {\Big
{)}}\ +\ {\frac {m\ v_{o}\ x_{o}}{{\hbar}}}$.\ \ \ \ \ (3.28)}
\end{center}
\par
Taking the eq. (3.28) in the eq. (3.12a) results:
\begin{center}
{$S(x,\ t)\ =\ {\frac {1}{{\hbar}}}\ {\int}_{o}^{t}\ dt'\ {\Big {(}}\
{\frac {1}{2}}\ m\ {\dot {q}}^{2}(t')\ +\ {\frac {m\ {\nu}}{2}}\
[{\frac {{\dot {a}}(t)}{a(t)}}\ -\ {\frac {{\nu}}{2}}]\ -\ V[q(t'),\
t']\ -\ {\frac {{\hbar}^{2}}{4\ m\ a^{2}(t')}}\ {\Big {)}}\ +$}
\end{center}
\begin{center}
{$+\ {\frac {m\ v_{o}\ x_{o}}{{\hbar}}}\ +\ {\frac {m\ {\dot
{q}}(t)}{{\hbar}}}\ [x\ -\ q(t)]\ +\ {\frac {m}{2\ {\hbar}}}\ {\Big
{[}}\ {\frac {{\dot {a}}(t)}{a(t)}}\ -\ {\frac {{\nu}}{2}}\ {\Big {]}}\
[x\ -\ q(t)]^{2}$\ .\ \ \ \ \ (3.29)}
\end{center}
\par
The above result permit us, finally, to obtain the wave packet
for the linearized Schuch-Chung-Hartmann equation along a classical
trajetory. Indeed, considering (2.2), (2.6), (3.1) and (3.29), we
get:\ [6]
\begin{center}
{${\psi}(x,\ t)\ =\ [2\ {\pi}\ a^{2}(t)]^{-\ 1/4}\ exp\ {\Big {[}}\ {\Big
{(}}\ {\frac {i\ m}{2\ {\hbar}}}\ [{\frac {{\dot
{a}}(t)}{a(t)}}\ -\ {\frac {{\nu}}{2}}]\ -\ {\frac {1}{4\ a^{2}(t)}}\
{\Big {)}}\ [x\ -\ q(t)]^{2}\ {\Big {]}}\ {\times}$}
\end{center}
\begin{center}
{${\times}\ exp\ {\Big {[}}\ {\frac {i\ m\ {\dot {q}}(t)}{{\hbar}}}\ [x\
-\ q(t)]\ +\ {\frac {i\ m\ v_{o}\ x_{o}}{{\hbar}}}\ {\Big {]}}\ {\times}$}
\end{center}
\begin{center}
{${\times}\ exp\ {\Big {[}}\ {\frac {i}{{\hbar}}}\ {\int}_{o}^{t}\ dt'\
{\Big {(}}\ {\frac {1}{2}}\ m\ {\dot {q}}^{2}(t')\ +\ {\frac {m\
{\nu}}{2}}\ [{\frac {{\dot {a}}(t)}{a(t)}}\ -\ {\frac {{\nu}}{2}}]\ -\
V[q(t'),\ t']\ -\ {\frac {{\hbar}^{2}}{4\ m\ a^{2}(t')}}\ {\Big {)}}\
{\Big {]}}$\ .\ \ \ \ \ (3.30)}
\end{center}
\par
Note that putting ${\nu}\ =\ 0$ into (3.30) we obtain the
quantum wave packet of the Schr\"{o}dinger equation with the potential
V(x,\ t).\ [7]
\par
4.\ {\bf The Feynman-de Broglie-Bohm Propagator of the Linearized
Schuch-Chung-Hartmann Equation along a Classical Trajetory}
\par
\vspace{0.2cm}
\par
4.1.\ {\bf Introduction}
\vspace{0.2cm}
\par
In 1948,\ [8] Feynman formulated the following principle of minimum
action for the Quantum Mechanics:
\begin{center}
{{\it The transition amplitude between the states ${\mid}\ a\ >$ and
${\mid}\ b\ >$ of a quantum-mechanical system is given by the sum of
the elementary contributions, one for each trajectory passing by
${\mid}\ a\ >$ at the time t$_{a}$ and by ${\mid}\ b\ >$ at the time
t$_{b}$. Each one of these contributions have the same modulus, but its
phase is the classical action S$_{c{\ell}}$ for each trajectory.}}
\end{center}
\par
This principle is represented by the following expression known as the
"Feynman propagator":
\begin{center}
{$K(b,\ a)\ =\ {\int}_{a}^{b}\ e^{{\frac {i}{{\hbar}}}\ S_{c{\ell}}(b,\
a)}\ D\ x(t)$\ ,\ \ \ \ \ (4.1)}
\end{center}
with:
\begin{center}
{$S_{c{\ell}}(b,\ a)\ =\ {\int}_{t_{a}}^{t_{b}}\ L\ (x,\ {\dot {x}},\
t)\ dt$\ ,\ \ \ \ \ (4.2)}
\end{center}
where $L(x,\ {\dot {x}},\ t)$ is the Lagrangean and $D\ x(t)$ is the
Feynman's Measurement. It indicates that we must perform the integration
taking into account all the ways connecting the states ${\mid}\ a\ >$
and ${\mid}\ b\ >$.
\par
Note that the integral which defines $K(b,\ a)$\ is called "path
integral" or "Feynman integral" and that the Schr\"{o}dinger
wavefunction ${\psi}(x,\ t)$ of any physical system is given by (we
indicate the initial position and initial time by $x_{o}$ and $t_{o}$,
respectively):\ [9]
\begin{center}
{${\psi}(x,\ t)\ =\ {\int}_{-\ {\infty}}^{+\ {\infty}}\ K(x,\ x_{o},\
t,\ t_{o})\ {\psi}(x_{o},\ t_{o})\ dx_{o}$\ ,\ \ \ \ \ (4.3)}
\end{center}
with the quantum causality condition:
\begin{center}
{${\lim\limits_{t,\ t_{o}\ {\to}\ 0}}\ K(x,\ x_{o},\ t,\ t_{o})\ =\
{\delta}(x\ -\ x_{o})$\ .\ \ \ \ \ (4.4)}
\end{center}
\vspace{0.2cm}
\par
4.2.\ {\bf Calculation of the Feynman-de Broglie-Bohm Propagator
for the Li\-nearized Schuch-Chung-Hartmann Equation along a Classical
Trajetory}
\vspace{0.2cm}
\par
According to Section 3, the wavefunction ${\psi}(x,\ t)$ that was
named wave packet of the of the linearized Schuch-Chung-Hartmann equation
along a classical trajetory, can be written as [see (3.30)]:
\begin{center}
{${\psi}(x,\ t)\ =\ [2\ {\pi}\ a^{2}(t)]^{-\ 1/4}\ exp\ {\Big {[}}\ {\Big
{(}}\ {\frac {i\ m}{2\ {\hbar}}}\ [{\frac {{\dot
{a}}(t)}{a(t)}}\ -\ {\frac {{\nu}}{2}}]\ -\ {\frac {1}{4\ a^{2}(t)}}\
{\Big {)}}\ [x\ -\ q(t)]^{2}\ {\Big {]}}\ {\times}$}
\end{center}
\begin{center}
{${\times}\ exp\ {\Big {[}}\ {\frac {i\ m\ {\dot {q}}(t)}{{\hbar}}}\ [x\
-\ q(t)]\ +\ {\frac {i\ m\ v_{o}\ x_{o}}{{\hbar}}}\ {\Big {]}}\ {\times}$}
\end{center}
\begin{center}
{${\times}\ exp\ {\Big {[}}\ {\frac {i}{{\hbar}}}\ {\int}_{o}^{t}\ dt'\
{\Big {(}}\ {\frac {1}{2}}\ m\ {\dot {q}}^{2}(t')\ +\ {\frac {m\
{\nu}}{2}}\ [{\frac {{\dot {a}}(t)}{a(t)}}\ -\ {\frac {{\nu}}{2}}]\ -\
V[q(t'),\ t']\ -\ {\frac {{\hbar}^{2}}{4\ m\ a^{2}(t')}}\ {\Big {)}}\
{\Big {]}}$\ .\ \ \ \ \ (4.5)}
\end{center}
where [see (3.24,25)]:
\begin{center}
{${\ddot {q}}(t)\ +\ {\nu}\ {\dot {q}}(t)\ +\ {\frac {1}{m}}\ V'[q(t),\
t]\ =\ 0$,\ \ \ \ \ (4.6)}
\end{center}
\begin{center}
{${\ddot {a}}(t)\ +\ a(t)\ {\Big {(}}\ {\frac {1}{m}}\ V"[q(t),\ t]\
\ -\ {\frac {{\nu}^{2}}{4}}\ {\Big {)}}\ =\ {\frac {{\hbar}^{2}}{4\
m^{2}\ a^{3}(t)}}$\ .\ \ \ \ \ (4.7)}
\end{center}
where the following initial conditions were obeyed [see (3.26a-d)]:
\begin{center}
{$q(0)\ =\ x_{o}\ ,\ \ \ {\dot {q}}(0)\ =\ v_{o}\ ,\ \ \ a(0)\
=\ a_{o}\ ,\ \ \ {\dot {a}}(0)\ =\ b_{o}$\ .\ \ \ \ \ \
(4.8a-d)}
\end{center}
\par
Therefore, considering (4.3), the Feynman-de Broglie-Bohm propagator
will be calculated using (4.5), in which we will put with no loss of
generality, $t_{o}\ =\ 0$. Thus:
\begin{center}
{${\psi}(x,\ t)\ =\ {\int}_{-\ {\infty}}^{+\ {\infty}}\ K(x,\ x_{o},\
t)\ {\psi}(x_{o},\ 0)\ dx_{o}$\ .\ \ \ \ \ (4.9)}
\end{center}
\par
Let us initially define the normalized quantity:
\begin{center}
{${\Phi}(v_{o},\ x,\ t)\ =\ (2\ {\pi}\ a_{o}^{2})^{1/4}\ {\psi}(v_{o},\
x,\ t)$\ ,\ \ \ \ \ (4.10)}
\end{center}
which satisfies the following completeness relation:\ [10]
\begin{center}
{${\int}_{-\ {\infty}}^{+\ {\infty}}\ dv_{o}\ {\Phi}^{*}(v_{o},\ x,\
t)\ {\Phi}(v_{o},\ x',\ t)\ =\ ({\frac {2\ {\pi}\ {\hbar}}{m}})\
{\delta}(x\ -\ x')$\ .\ \ \ \ \ (4.11)}
\end{center}
\par
Taking the eqs. (2.2,6), we have:
\begin{center}
{${\psi}^{*}(x,\ t)\ {\psi}(x,\ t)\ =\ {\phi}^{2}\ =\ {\rho}(x,\ t)$\
.\ \ \ \ \ (4.12)}
\end{center}
\par
Now, using the eqs. (4.10,12), we get:
\begin{center}
{${\Phi}^{*}(v_{o},\ x,\ t)\ {\psi}(v_{o},\ x,\ t)\ =$}
\end{center}
\begin{center}
{$=\ (2\ {\pi}\ a_{o}^{2})^{1/4}\ {\psi}^{*}(v_{o},\ x,\ t)\
{\psi}(v_{o},\ x,\ t)\ =\ (2\ {\pi}\ a_{o}^{2})^{1/4}\ {\rho}(v_{o},\
x,\ t)\ \ \ {\to}$}
\end{center}
\begin{center}
{${\rho}(v_{o},\ x,\ t)\ =\ (2\ {\pi}\ a_{o}^{2})^{-\ 1/4}\
{\Phi}^{*}(v_{o},\ x,\ t)\ {\psi}(v_{o},\ x,\ t)$\ .\ \ \ \ \
(4.13)}
\end{center}
\par
On the other side, substituting (4.13) into (3.6b), integrating the
result and using (3.1) and (4.10) results [remembering
that ${\psi}^{*}\ {\psi}({\pm}\ {\infty})\ \ \ {\to}\ \ \ 0$]:
\begin{center}
{${\frac {{\partial}({\Phi}^{*}\ {\psi})}{{\partial}t}}\ +\ {\frac
{{\partial}({\Phi}^{*}\ {\psi}\ {\vartheta}_{qu})}{{\partial}x}}\ =\ 0\
\ \ {\to}$}
\end{center}
\begin{center}
{${\frac {{\partial}}{{\partial}t}}\ {\int}_{-\ {\infty}}^{+\ {\infty}}\
dx\ {\Phi}^{*}\ {\psi}\ +\ ({\Phi}^{*}\ {\psi}\
{\vartheta}_{qu}){\mid}_{-\ {\infty}}^{+\ {\infty}}\ =$}
\end{center}
\begin{center}
{$=\ {\frac {{\partial}}{{\partial}t}}\ {\int}_{-\ {\infty}}^{+\ {\infty}}\
dx\ {\Phi}^{*}\ {\psi}\ +\ (2\ {\pi}\ a_{o}^{2})^{1/4}\ ({\psi}^{*}\
{\psi}\ {\vartheta}_{qu}){\mid}_{-\ {\infty}}^{+\ {\infty}}\ =\ 0\ \ \
{\to}$}
\end{center}
\begin{center}
{${\frac {{\partial}}{{\partial}t}}\ {\int}_{-\ {\infty}}^{+\
{\infty}}\ dx\ {\Phi}^{*}\ {\psi}\ =\ 0$\ .\ \ \ \ \ (4.14)}
\end{center}
\par
The eq. (4.14) shows that the integration is time independent.
Consequently:
\begin{center}
{${\int}_{-\ {\infty}}^{+\ {\infty}}\ dx'\ {\Phi}^{*}(v_{o},\ x',\ t)\
{\psi}(x',\ t)\ =\ {\int}_{-\ {\infty}}^{+\ {\infty}}\ dx_{o}\
{\Phi}^{*}(v_{o},\ x_{o},\ 0)\ {\psi}(x_{o},\ 0)$\ .\ \ \ \ \ (4.15)}
\end{center}
\par
Multiplying (4.15) by ${\Phi}(v_{o},\ x,\ t)$ and integrating over
$v_{o}$ and using (4.11), we obtain [remembering
that ${\int}_{-\ {\infty}}^{+\ {\infty}}\ dx'\ f(x')\ {\delta}(x' -\
x)\ = f(x)$]:
\begin{center}
{${\int}_{-\ {\infty}}^{+\ {\infty}}\ {\int}_{-\ {\infty}}^{+\
{\infty}}\ dv_{o}\ dx'\ {\Phi}(v_{o},\ x,\ t)\ {\Phi}^{*}(v_{o},\ x',\ t)\
{\psi}(x',\ t)$\ =}
\end{center}
\begin{center}
{=\ ${\int}_{-\ {\infty}}^{+\ {\infty}}\ {\int}_{-\
{\infty}}^{+\ {\infty}}\ dv_{o}\ dx_{o}\ {\Phi}(v_{o},\ x,\ t)\
{\Phi}^{*}(v_{o},\ x_{o},\ 0)\ {\psi}(x_{o},\ 0)\ \ \ {\to}$}
\end{center}
\begin{center}
{${\int}_{-\ {\infty}}^{+\ {\infty}}\ dx'\ ({\frac {2\ {\pi}\
{\hbar}}{m}})\ {\delta}(x'\ -\ x)\ {\psi}(x',\ t)\ =\ ({\frac {2\ {\pi}\
{\hbar}}{m}})\ {\psi}(x,\ t)$\ =}
\end{center}
\begin{center}
{=\ ${\int}_{-\ {\infty}}^{+\ {\infty}}\ {\int}_{-\ {\infty}}^{+\
{\infty}}\ dv_{o}\ dx_{o}\ {\Phi}(v_{o},\ x,\ t)\ {\Phi}^{*}(v_{o},\
x_{o},\ 0)\ {\psi}(x_{o},\ 0)\ \ \ {\to}$}
\end{center}
\begin{center}
{${\psi}(x,\ t)\ =\ {\int}_{-\ {\infty}}^{+\ {\infty}}\ {\Big {[}}\
({\frac {m}{2\ {\pi}\ {\hbar}}})\ {\int}_{-\ {\infty}}^{+\ {\infty}}\
dv_{o}\ {\Phi}(v_{o},\ x,\ t)\ {\times}$}
\end{center}
\begin{center}
{${\times}\ {\Phi}^{*}(v_{o},\ x_{o},\ 0)\ {\Big {]}}\ {\psi}(x_{o},\
0)\ dx_{o}$\ .\ \ \ \ \ (4.16)}
\end{center}
\par
Comparing (4.9) and (4.16), we have:
\begin{center}
{$K(x,\ x_{o},\ t)\ =\ {\frac {m}{2\ {\pi}\ {\hbar}}}\ {\int}_{-\
{\infty}}^{+\ {\infty}}\ dv_{o}\ {\Phi}(v_{o},\ x,\ t)\
{\Phi}^{*}(v_{o},\ x_{o},\ 0)$\ .\ \ \ \ \ (4.17)}
\end{center}
\par
Substituting (4.5) and (4.10) into (4.17), we finally obtain the
Feynman-de Broglie-Bohm Propagator of the linearized
Schuch-Chung-Hartmann equation along a classical trajetory [remembering
that ${\Phi}^{*}(v_{o},\ x_{o},\ 0)\ =\ exp\ (-\ {\frac {i\ m\ v_{o}\
x_{o}}{{\hbar}}})$]:
\begin{center}
{$K(x,\ x_{o};\ t)\ =\ {\frac {m}{2\ {\pi}\ {\hbar}}}\ {\int}_{-\
{\infty}}^{+\ {\infty}}\ dv_{o}\ {\sqrt {{\frac
{a_{o}}{a(t)}}}}\ {\times}$}
\end{center}
\begin{center}
{${\times}\ exp\ {\Big {[}}\ {\Big
{(}}\ {\frac {i\ m}{2\ {\hbar}}}\ [{\frac {{\dot
{a}}(t)}{a(t)}}\ -\ {\frac {{\nu}}{2}}]\ -\ {\frac {1}{4\ a^{2}(t)}}\
{\Big {)}}\ [x\ -\ q(t)]^{2}\ +\ {\frac {i\ m\ {\dot
{q}}(t)}{{\hbar}}}\ [x\ -\ q(t)]\ {\Big {]}}\ {\times}$}
\end{center}
\begin{center}
{${\times}\ exp\ {\Big {[}}\ {\frac {i}{{\hbar}}}\
{\int}_{o}^{t}\ dt'\ {\Big {(}}\ {\frac {1}{2}}\ m\ {\dot
{q}}^{2}(t')\ +\ {\frac {m\ {\nu}}{2}}\ [{\frac {{\dot {a}}(t)}{a(t)}}\
-\ {\frac {{\nu}}{2}}]\ -\ V[q(t'),\ t']\ -\ {\frac {{\hbar}^{2}}{4\ m\
a^{2}(t')}}\ {\Big {)}}\ {\Big {]}}$\ ,\ \ \ (4.18)}
\end{center}
where $q(t)$ and $a(t)$ are solutions of the (4.6,\ 7) differential
equations.
\par
Finally, it is important to note that putting ${\nu}\ =\ 0$ and
V[q(t'),\ t']\ =\ 0 into (4.6), (4.7) and (4.18) we obtain the free
particle Feynman propagator.\ [1,\ 9]
\par
\begin{center}
{{\bf NOTES AND REFERENCES}}
\end{center}
\par
1.\ BASSALO, J. M. F., ALENCAR, P. T. S., CATTANI, M. S. D. e
NASSAR, A. B. {\it T\'opicos da Mec\^anica Qu\^antica de de
Broglie-Bohm}, EDUFPA (2003).
\par
2.\ SCHUCH, D., CHUNG, K. M. and HARTMANN, H. {\it Journal of
Mathematical Physics 24}, p. 1652 (1983); {\it Journal of
Mathematical Physics 25}, p. 3086 (1984); {\it Berichte der
Bunsen-Gesellschaft f\"{u}r Physikalische Chemie 89}, p. 589
(1985).
\par
3.\ MADELUNG, E. {\it Zeitschrift f\"{u}r Physik 40}, 322 (1926).
\par
4.\ BOHM, D. {\it Physical Review 85}, 166 (1952).
\par
5.\ BASSALO, J. M. F., ALENCAR, P. T. S., SILVA, D. G., NASSAR, A. B.
and CATTANI, M. {\it arXiv:0905.4280v1}\ [quant-ph]\ 26\ May\
2009; -----. {\it arXiv:1004.1416v1}\ [quant-ph]\ 10\ April\
2010; -----. {\it arXiv:1006.1868v1}\ [quant-ph]\ 9\ June\
2010.
\par
6. NASSAR, A. B., BASSALO, J. M. F., ALENCAR, P. T. S., CANCELA, L. S.
G. and CATTANI, M. {\it Physical Review 56E}, 1230 (1997).
\par
7.\ NASSAR, A. B. {\it Journal of Mathematics Physics 27}, 2949 (1986).
\par
8.\ FEYNMAN, R. P. {\it Reviews of Modern Physics 20}, 367 (1948).
\par
9.\ FEYNMAN, R. P. and HIBBS, A. R. {\it Quantum Mechanics and Path
Integrals}, McGraw-Hill Book Company (1965).
\par
10.\ BERNSTEIN, I. B. {\it Physical Review A32}, 1 (1985).
\end{document}